\documentclass[12pt,preprint]{aastex}
\shorttitle{Exact reconstruction of 4- and 5-dimensional inflationary potential}
\shortauthors{L.A. Popa et al. }
\begin{document}

\title{From WMAP to {\sc Planck}: Exact reconstruction of 4- and 5-dimensional inflationary potential from high precision CMB measurements}
\author{L.A. Popa\altaffilmark{1}, N. Mandolesi \altaffilmark{2},
 A. Caramete\altaffilmark{1}, C. Burigana\altaffilmark{2}}
\affil{$^1 \, $Institutul de \c{S}tiin\c{t}e Spa\c{t}iale Bucure\c{s}ti-M\u{a}gurele, Ro-077125 Rom\^{a}nia}
\affil{$^2 \,$INAF/IASF, Istituto di Astrofisica Spaziale e Fisica
Cosmica Bologna, I-40129, Italia}

\email{lpopa@venus.nipne.ro }
%Article-id: 0907.5558, Article password: 6pcg8 (access still password restricted)

\begin{abstract}
We make a more general determination of the inflationary observables
in the standard 4-D and 5-D single-field inflationary scenarios,
by the exact reconstruction of the dynamics of the inflation potential during the
observable inflation with minimal number of assumptions:
the computation does not assume the slow-roll approximation and  is valid in all regimes
if the field is monotonically rolling down its potential.

We address higher-order effects in the standard and braneworld single-field
inflation scenarios by fitting the Hubbble expansion rate
and subsequently the inflationary potential directly
to WMAP5+SN+BAO and Planck-like simulated datasets. \\
Making use of the {\em Hamilton-Jacobi} formalism developed for the 5-D single-field inflation model,
we compute the scale dependence of the amplitudes of the scalar
and tensor perturbations by integrating the exact mode equation.
The solutions in 4-D and 5-D inflation scenarios differ through the dynamics
of the background scalar field and the number
of {\it e}-folds assumed to be compatible with the observational window of inflation.

We analyze the implications of the theoretical uncertainty in the determination of the reheating
temperature after inflation on the observable predictions of inflation
and evaluate its  impact on the degeneracy of the standard 
inflation consistency relation.

We find that the detection of tensor perturbations and
the theoretical uncertainties in the inflationary observable represent
a significant challenge for the future {\sc Planck} CMB measurements:
distinguishing  between the observational signatures of the standard and
braneworld single-field inflation scenarios.

\vspace{0.7cm}
This work have been done in the frame of {\sc Planck} Core Team activities.
\end{abstract}


\section{Introduction}

The primary goal of particle cosmology is to obtain a
concordant  description of the origin and early evolution 
of the Universe,
consistent with both unified field theory and astrophysical 
and cosmological
measurements. \\
Inflation is the most simple and robust theory able 
to explain the astrophysical and cosmological observations,
providing at the same time self-consistent primordial 
initial conditions \cite{Staro80,Guth81, Sato81,
Albercht82,Linde83,Linde83} and the mechanisms 
for quantum generation of the scalar and tensor
perturbations 
\cite{Muk81,Haw82,Staro82,Guth82,Bard83,Abo84}.\\
In the simplest class of inflationary models, inflation is driven by a single
scalar field $\phi$ (inflaton) with some potential
$V(\phi)$, minimally coupled to  Einstein gravity. The perturbations are predicted
to be adiabatic, nearly scale-invariant and Gaussian distributed,
resulting in an effectively flat Universe.
At the leading-order in slow-roll approximation
\cite{Stein84,Salopek90,Pars94} the amplitudes of scalar
and tensor perturbations on a specified comoving wavenumber $k$,
are related through the consistency equation:
\begin{equation}
\label{con0}
\frac {A^2_T  }  { A^2_S  } =-\frac{n_T}{2}  \,,
\end{equation}
where: $A^2_S \propto k^{n_S-1}$ and $A^2_T \propto k^{n_T}$ are
the amplitudes of  scalar and tensor perturbations respectively and $n_S$ and $n_T$ are their tilts.
The consistency equation may be regarded as an independent test of single-field 
inflationary models as it does not depend on the specific functional form of the inflation
potential. \\
Recent WMAP 5-year CMB measurements, alone \cite{Dunkeley09} or
complemented with other cosmological datasets \cite{Komatsu09},
support  the standard inflationary predictions of a nearly flat Universe with adiabatic initial density perturbations. \\
In particular, the detected anti-correlations between temperature and E-mode polarization
anisotropy on degree scales \cite{Nolta09} provide strong evidence for correlation on
length scales beyond the Hubble radius.

Despite the successes of inflationary cosmology, recent proposals in theoretical
physics motivated by the developments in superstring and M-theory \cite{Hora}, suggest that our four-dimensional Universe could lie on a brane embedded in higher-dimensional space-time
(see e.g. Rubakov 2001, Maartens 2004 and references therein). \\
In particular, in the type II Randall-Sundrum model (RSII) \cite{Randall99a,Randall99b}
our four-dimensional (4-D) Universe is a brane with positive tension
$\lambda$ embed in a five-dimensional (5-D) anti-de Sitter space-time (AdS$_5$).
At sufficiently low energies ($\rho << \lambda$) the standard cosmic behavior is recovered and the primordial nucleosynthesis constraint is satisfied, provided that $\lambda\geq$(1 MeV)$^4$. \\
The simplest way to realize inflation in RSII model is to have a single scalar field
confined to the brane and only gravity in the bulk \cite{MWBH02}.
In this case the Fiedmann equation is modified at so that Hubble parameter $H \propto \rho$ rather than $H \propto \rho^{1/2}$ as in 4-D case,
leading to significantly modifications of the amplitudes and scale dependencies of scalar and tensor perturbations \cite{Bine99}.

The observational constraints on inflationary parameters in 5-D scenario, made in general by using
the slow-roll (SR) approximation in the high energy regime ($\rho >> \lambda$),
show that to the leading order in SR the consistency equation
has precisely the same form as in the standard 4-D scenario,
the relationship between inflationary observables being independent on the brane tension
\cite{LidT04,SeT05}.\\
The degeneracy of consistency equation was associated with the fact that 5-D inflationary observables smoothly approach their 4-D counterparts  as the brane decouples from the bulk approaching
the low energy regime ($\rho << \lambda$).
The main assumption made by these works is that the back-reaction due to metric perturbation
in the bulk can be neglected. This assumption is valid to the leading-order in slow-roll
approximation, as the coupling between inflation fluctuations and metric perturbation vanishes.\\
Recently it was shown \cite{Koya04,Koya05a,Koya05b} that the sub-horizon
inflation fluctuations on the brane excite an
infinite ladder of Klauza-Klein modes of the bulk metric
perturbations to second-order in slow-roll parameters.
If the back-reaction is take into account, the amplitude of the scalar perturbations receives second-order slow-roll corrections in addition to Stewart-Lyth corrections \cite{Stew93}, of the same order of magnitude \cite{Koya08}. The degeneracy of consistency equations  does not hold
when the second-order corrections in SR expansion for perturbations are included \cite{Calg03,Calg04,RamL04,SeT05}.

One of the most anticipated results of forthcoming high precision CMB
experiments is probing the physics of inflation and in particular the reconstruction
of the inflation potential. In order to have a robust interpretation of upcoming observations
it is imperative to understand how the reconstruction process
may be affected by the degeneracy of the inflationary observables.
In this paper we aim to make a more general determination
of the inflationary observables in 4-D and 5-D inflationary scenarios,
by exact reconstruction of the dynamics of the inflation potential during the
observable inflation with minimal number of assumptions \cite{Julien08, Julien09}. \\
Taking the advantage of the formalism developed for the standard single-field
inflation \cite{Peiris03,Peiris06a,Peiris06b,Martin06,Julien08,Lidsey08}, we carry out similar
calculations for 5-D inflation models by fitting the Hubbble function, $H(\phi)$,
and subsequently the inflationary potential, $V(\phi)$, directly
to WMAP 5-year data \cite{Dunkeley09,Komatsu09} complemented  with
geometric probes  from the Type Ia supernovae (SN) distance-redshift relation and
the baryon acoustic oscillations (BAO) measurements and Planck-like CMB anisotropy simulated data. \\
Our specific goal is to address higher-order effects in the standard and braneworld single-field
inflation models and to analyze the sensitivity of the present and
future CMB temperature and polarization measurements to discriminate between them. 
 
The paper is organized as follows. In Section~2 we review the Hamilton-Jacobi
formalisms for 4-D and 5-D sigle-field inflation models.
In Section~3 we compute the scalar and tensor perturbation spectra
for standard  and  braneworld  single-field infation models by using the
exact mode equation.
In Section~4 we present the implementation of  the Markov Chain Monte-Carlo methodology
and describe the datasets involved in our analysis.
Section~5 is dedicated to the analysis and the interpretation of our results: we present the derived bounds on the inflationary parameters, Hubble Slow-Roll parameters and the magnitude, slope and curvature of the infationary potentials obtained from the fits of 4-D and 5-D single-field inflation models to our datasets and analyze the possibility do disentangle between standard and braneworld scenarios by using the future {\sc Planck} high precision CMB measurements.
In Section~6 we draw our conclusions.

Throughout the paper $m_4$ and $m_5$ denote the corresponding 4-D and 5-D
Planck mass scales and we have set G$m^2_4= \hbar=c=1$.
Also, we denote by dot the derivative with respect to the time and by prime the derivative with
respect to the scalar field.

\section{The four- and five-dimensional Hamilton-Jacobi formalism}
\subsection{The 4-D single-field inflation case}

The Hubble Slow-Roll (HSR) formalism for the standard
single-field inflation was set down in detail by
Liddle, Parsons \& Barrow (1994).\\
The Friedmann equation in zero-curvature Universe is given by:
\begin{equation}
\label{Fri4D}
H_{4D}^2= \frac{8 \pi}{3 m^2_4}\rho \,,
\end{equation}
where: $H \equiv {\dot a}/a$ is the Hubble parameter, $a$ is the cosmological scale factor,
$\rho=V+{\dot \phi}^2/2$ is the total energy density, where $V(\phi)$
and ${\dot \phi}^2/2$ are the potential and kinetic  energy density terms respectively.
Since the dark energy contribution is strongly suppressed by the exponential expansion during inflation \cite{MWBH02,Sasaki01}, we set to zero the dark energy term
in the above equation.\\
The equation of motion for the scalar field is given by:
\begin{equation}
\label{eqphi}
{\ddot \phi}+3H {\dot \phi}=-V' \,.
\end{equation}
Eqs. (\ref{Fri4D}) and (\ref{eqphi}) can be written in the {\em Hamilton-Jacobi} form,
allowing to consider inflation in terms of  $H(\phi)$
rather than $V(\phi)$ \cite{Pars94,Kinney02,Easther03,Peiris03,Kinney04}:
\begin{eqnarray}
\label{Ham1}
H'(\phi) a'(\phi) & = & -\frac{4 \pi}{m^2_4} H(\phi) a(\phi) \,, \\
\label{Ham2}
{\dot \phi } & = &  -\frac{m^2_4}{4 \pi}H'(\phi) \,, \\
\label{Ham3}
\left[ H'(\phi)\right ]^2 - \frac{12 \pi}{m^2_4} H^2(\phi) &  = &
-\frac{32 \pi^2}{m^4_4}V(\phi) \,.
\end{eqnarray}
For any value of $H(\phi$) Eq. (\ref{Ham3}) can be used to find $V(\phi)$ while Eqs. (\ref{Ham1}) and (\ref{Ham2}) allow to convert $\phi$-dependence into time-dependence.\\
In the standard 4-D inflation
the first three HSR parameters are given \cite{Pars94}:
\begin{eqnarray}
\label{eps4D}
\epsilon_H & = & \frac{m^2_4}{4 \pi} \frac{H'^2(\phi)}{H^2(\phi)} \,, \\
\label{eta4D}
\eta_H (\phi) & = &
\frac{m^2_4}{4 \pi} \frac{H''(\phi)}{H(\phi)} \,,\\
\label{zeta4D}
\zeta^2_H (\phi) & = &
\frac{m^4_4}{16 \pi^2} \frac{H'(\phi)\,H'''(\phi)}{H(\phi)} \,.
\end{eqnarray}
The dependence of $V(\phi)$ on $H(\phi)$ can be obtain by substituting $\epsilon_H$ into Eq. (\ref{Ham3}) leading to:
\begin{equation}
\frac{8 \pi}{3 m^2_4} V(\phi)=H^2(\phi)\left[1-\frac{1}{3}\epsilon_H(\phi)\right]\,.
\end{equation}
The HSR formalism ensures that the condition for inflation to occur is precisely $\epsilon_H < 1$ and inflation ends exactly when $\epsilon_H=1$.

\subsection{The 5-D single-field inflation case}

In the 5-D inflation case the Eq.(\ref{Fri4D}) receives an additional
term quadratic in energy density:
\begin{equation}
\label{Fri5D}
H^2_{5D}=\frac{8 \pi}{3m^2_4}\rho
\left( 1+\frac{\rho}{2 \lambda} \right) \,,
\end{equation}
where: $\rho=V+{\dot \phi/2}$ is the total energy density and $\lambda$ is the brane tension. 
 The scalar field $\phi$ is assumed to obey the same equation
of motion as in 4-D standard inflation as given by Eq.(\ref{eqphi}). 
In the low-energy regime ($\rho << \lambda$) the quadratic term in Eq.(\ref{Fri5D}) 
can be neglected and one recover the behavior of the 4-D standard cosmology.
In high-energy regime ($\rho >> \lambda$) the deviation from the standard expansion changes the amplitudes and scale-dependence of cosmological perturbations. \\
Hereafter we will make use of the approach developed by
Hawkins and Lidsey who derived a general formalism for 5-D inflation case
valid in all regimes, having many of the properties of the {\em Hamilton-Jacobi}
formalism  in 4-D standard inflation \cite{HawL01,HawL03}.
They defined a quantity $y(\phi)$ with the same role as $H(\phi)$ in the case of 4-D standard inflation:
\begin{eqnarray}
\label{yphi}
y^2(\phi)  =  \frac{\rho/2 \lambda}{1+\rho/2 \lambda} \,,
\end{eqnarray}
with the inverse relation given by:
\begin{eqnarray}
\label{rhophi}
\rho  =  \frac{2 \lambda y^2(\phi)}{1-y^2(\phi)} \,.
\end{eqnarray}
In terms of $y(\phi)$ the Friedmann equation (\ref{Fri5D})  reads as:
\begin{equation}
\label{Friy}
H^2_{5D}(y)= \frac{16 \pi \lambda}{3 m^2_4}
\frac{y^2(\phi)}{(1-y^2(\phi))^2} \,,
\end{equation}
where the restriction $y^2<1$ is imposed,
implying that $y(\phi)$ is proportional to $H(\phi)$
in the low-energy limit, $y \rightarrow 0$ ($\rho/\lambda \rightarrow 0$).\\
The Hamilton-Jacobi equations, analogues to Eqs. (\ref{Ham1}) - (\ref{Ham3}) for
 4-D standard inflation  are given by \cite{HawL03,RamL04}:
\begin{eqnarray}
\label{Ham51}
y'(\phi)a'(\phi) & = & -\frac{4 \pi}{m^2_4}y(\phi)a(\phi) \,, \\
\label{Ham52}
{\dot \phi} & = &-\left( \frac{\lambda m^2_4}{3 \pi} \right)^{1/2} \frac{y'}{1-y^2} \,, \\
\label{Ham53}
H'(\phi) & = &-\frac{4 \pi}{m^2_4}\frac{(1+y^2)}{(1-y^2)} \, {\dot \phi} \,,
\end{eqnarray}
and the dependence of $V(\phi)$ on $y(\phi)$ can be obtained by combining Eqs.(\ref{Ham52}) and (\ref{Ham53}) leading to:
\begin{eqnarray}
\label{V5}
V(\phi)=\frac{2 \lambda y^2}{1-y^2}-\frac{\lambda m^2_4}{6 \pi}
\left( \frac{y'}{1-y^2} \right) \,.
\end{eqnarray}
The first three HSR parameters in terms of  $y(\phi)$ reads as \cite{RamL04}:
\begin{eqnarray}
\label{eps5D}
\epsilon_H & = & \left(\frac{\lambda m^2_4}{3 \pi} \right)^{1/2} \, \frac{y}{(1+y^2)}\frac{H'^2}{H^3} \,, \\
\label{eta5D}
\eta_H & = & \left(\frac{\lambda m^2_4}{3 \pi}\right)^{1/2}\left[ \frac{y}{(1+y^2)}\frac{H''}{H^2}- \frac{4y^3}{(1+y^2)^3}\frac{H'^2}{H^3} \right]  \, ,\\
\label{zeta5D}
\xi^2_H & = & = \frac{{\ddot \phi}}{H^2{\dot \phi}}-\eta^2_H \,.
\end{eqnarray}
The above definitions of HSR parameters are valid in all regimes,
generalizing the previous ones, preserving at the same time
many of the inflation key properties:
they are obtained by demanding  the condition for inflation
to occur precisely for $\epsilon_H < 1$ and to end exactly when $\epsilon_H=1$.
Also, they are preserving the lowest-order
slow-roll definitions of the scalar spectral index, $n_S=1-4\epsilon_H+2\eta_H$,
and of its running,
$dn_S/d{\rm ln}k= 5\epsilon_H\eta_H -4 \epsilon^2_H-2 \xi^2_H $.

\section{The four- and five-dimensional exact mode equation}

The scale dependence of the amplitudes of the scalar (S) and tensor (T)
perturbations can be exactly obtained  by integrating the mode  equation  \cite{Muk85,Muk89}:
\begin{eqnarray}
\label{Muk}
u_k''+ \left(k^2- \frac{z''_{S,T}}{z_{S,T}} \right)u_k  =  0 \,,
\end{eqnarray}
where primes denote the second derivatives with respect to the conformal time.\\
The numerical evaluation of the spectra involves solving Eq.(\ref{Muk})
for each value of the wavenumber $k$, the evolution of $|u_k|/z_{S,T}$ to a constant value
defining the observable power spectra ${\cal P}_{S,T}$. 
The solutions differ through the evolution of the background scalar field
and the prior on the number of {\it e}-folds assumed to be compatible with
the observational window of inflation.

\subsection{The 4-D single-field inflation case}
We compute the amplitudes of scalar and tensor perturbations
by using the standard inflation numerical module
from Lesgourgues et al. (2008). 
For each wavenumber $k$ in a given range
the code integrates Eq.(\ref{Muk})
in an observational inflationary window corresponding to a number of $\Delta N$
 {\it e}-folds, imposing that
$k$ grows monotonically to the wavenumber $k_*$
that leaves the Hubble radius when $\phi=\phi_*$,
eliminating at the same time the models violating the condition
for inflation ($\epsilon_H < 1)$. \\
For the purpose of present analysis we reconstruct the Hubble expansion rate $H(\phi-\phi_*)$
from the data by using the Taylor expansion up to the cubic term:
\begin{eqnarray}
H(\phi-\phi_*)=H_*+H'_*(\phi-\phi_*)+\frac{1}{2}H''_*(\phi-\phi_*)^2 +
\frac{1}{6}H'''_*(\phi-\phi_*)^3 \,,
\end{eqnarray}
equivalent to keeping the first three HSR parameters.
We consider wavenumbers in the range [$5 \times 10^{-6}-5$]~Mpc$^{-1}$
needed to numerically derive the CMB angular power spectra
and the Hubble crossing  scale $k_*=0.01$Mpc$^{-1}$.\\
The analysis however depends on the prior on the interval over which
the dynamics of the background field is tracked.
Actually, the standard inflation numerical module integrates Eq.(\ref{Muk})
from the time at which
$k/aH=50$ until $d {\rm ln} {\cal P}_{S,T}/d{\rm ln} a < 3  \times 10^{-3}$. 
This choice ensures that inflation started
enough  time before the observational range and ends ($\epsilon_H=1$)
enough time after the smallest observable
scale leaves the Hubble horizon scale $k_*=0.01$ Mpc$^{-1}$,
leading at the same time to an accuracy of $\sim 0.1\%$ in final power spectra amplitudes, that is smaller than the expected sensitivity of CMB data \cite{Julien09}. 
For these reasons we choose to keep this time integration window for our computation. \\
The power spectra of scalar and tensor perturbations are obtained as \cite{Stew93,Copeland94}:
\begin{eqnarray}
\label{As_e}
{\cal P}_S (k)= \frac{k^3}{2 \pi^2} \, \left( \frac{H }{ {\dot \phi} }\right)^2 \,\frac{|u_k|^2}{a^2}  \, ,
\hspace{0.8cm}
\label{At_e}
{\cal P}_T (k)= \frac{16k^3}{\pi m^2_4} \,  \frac{|u_k|^2}{a^2}  \,,
\end{eqnarray}
where $z_S=a{\dot \phi}/H$ for scalars, $z_T=a$ for tensors and
the temporal evolution of the scalar field is given by Eq.(\ref{Ham2}).

\subsection{The 5-D single-field inflation case}
There is supporting evidence for the use of the exact mode Eq.(\ref{Muk})
in braneworld context, because its derivation does not involve the Friedmann equation
\cite{LiddleLyth00}.\\
We modify the standard inflation module from Lesgourgues et al. (2008)
to compute the  power spectra of scalar and tensor perturbations
for the single-field braneworld inflation. \\
As in the case of 4-D standard inflation, the Hubble expansion rate $H(\phi-\phi_*)$ is
obtained from the data by the Taylor expansion up to the cubic term
the neighborhood of the pivot scale $k_*$=0.01 Mpc$^{-1}$. \\
For each wavenumbers $k$ in the range [$5 \times 10^{-6}-5$]~Mpc$^{-1}$ we
integrate Eq.(\ref{Muk}) keeping the same time integration window as
in the 4-D inflation case. In this way, the identical scales (wavenumbers) encompass
the same number of  {\it e}-folds at Hubble radius crossing $k_*$ in both  4-D
and 5-D inflationary scenarios ensuring the same accuracy in the reconstruction of the inflationary potential \cite{Kinney02}.
As in the 4-D case we impose the condition that each mode
$k$ grows monotonically to the wavenumber $k_*$ and we
eliminate those models violating the condition for inflation ($\epsilon_H < 1)$. \\
Taking $z_S=aH/{\dot \phi}$ for scalars, $z_T=a$ for tensors and the evolution of the scalar field given by Eq.(\ref{Ham52}), the power spectra of scalar and tensor perturbations are then obtained as \cite{RamL04,Koya08}:
\begin{eqnarray}
\label{As_bw}
{\cal P}_S (k)= \frac{ k^3}{25 \pi^2} \, \left( \frac{H }{ {\dot \phi} }\right)^2 \,\frac{|u_k|^2}{a^2}  \,[1+{\cal K}(\beta^2)]\,,
\hspace{0.8cm}{\cal P}_T (k)=
\frac {16 k^3} { \pi  m^2_4 }  \frac{ |u_k|^2}{a^2} \,.
\end{eqnarray}
The correction ${\cal K}(\beta^2)$ to ${\cal P}_S(k)$ is solely due to
the coupling between the
inflation field and the bulk metric perturbations \cite{Koya05a,Koya05b,Koya08}. 
For each wavenumber $k$ we obtained ${\cal K}(\beta^2)$ by numerical computation
\cite{Koya08}, taking $\beta^2$ given by \cite{Koya08}:
\begin{equation}
\beta^2=\frac{1}{3}\epsilon_H \left[1+\left(\frac{\mu}{H}\right)^2\right]^{-1/2} \,.
\end{equation}
One should note that, although  the power spectra of the tensor perturbations
in 4-D and 5-D inflation have the same form, they differ through their dependencies
on the cosmological scale factor $a(\phi)$ and on the conformal time: $\eta(a)=\int da/a^2H$. \\
Defining the amplitudes of scalar and tensor power spectra as \cite{Copeland94}\footnote{The normalization of ${\cal P}_S(k)$ ensures that $A_S$ coincides precisely with the density contrast $\delta_H$
at Hubble radius crossing as defined by Liddle and Lyth \cite{LiddleLyth00}.
The normalization of  ${\cal P}_T(k)$ is then chosen so that $\epsilon_H=A^2_T/A^2_S$.}:
 $A_S(k)=2{\cal P}^{1/2}_S(k)/5$ and
$A_T(k)={\cal P}^{1/2}_T(k)/10$,
the scalar and tensor spectral indexes $n_{S,T}$ and the running of scalar tilt
$\alpha_S$ at the Hubble radius crossing $k=aH$ are defined as usual by:
\begin{eqnarray}
\label{inflpar_e}
n_S-1 & \equiv &  \left. \frac{d{\rm ln}A^2_S}{d{\rm ln}k}\right|_{k=aH}
\hspace{0.8cm}
n_T \equiv  \left. \frac{d{\rm ln}A^2_T}{d{\rm ln}k} \right|_{k=aH} \,
\hspace{0.8cm}
\alpha_S \equiv \left. \frac{dn_S}{d{\rm ln}k}\right|_{k=aH} \,.
\end{eqnarray}

\section{The Markov Chain Monte Carlo methodology}

We use the Markov Chain Monte Carlo (MCMC) technique to reconstruct the inflationary
potential and to derive constraints on the inflationary observables in the 4-D inflation and 5-D single-field inflation models by using 
the WMAP 5-year data \cite{Dunkeley09,Komatsu09} complemented  with
geometric probes  from the Type Ia supernovae (SN) distance-redshift relation and the baryon acoustic oscillations (BAO).\\
The SN distance-redshift relation has been studied in detail in the recent
unified analysis of the published heterogeneous SN data sets -
the Union Compilation08  \cite{Kowalski08}.\\
The BAO in the distribution
of galaxies  are extracted from the Sloan Digital Sky Surveys (SDSS) and Two Degree Field Galaxy Redshidt Survey (2DFGRS) \cite{Percival07}.
The CMB, SN and BAO data (WMAP5+SN+BAO) are combined by multiplying the likelihoods. 
We decided to use these measurements especially because we are testing models deviating from  the  standard Friedmann expansion.
These datasets properly enables us to account for any shift of the CMB
angular diameter distance  and of the expansion rate of the Universe.

For the forecast from {\sc Planck}-like simulated data we use the CMB temperature (T) and polarization (P) power spectra of our fiducial cosmological model and the expected  experimental characteristics  of the {\sc Planck} frequency channels given in Table~1 \cite{Mandolesi_etal_M1,Blue}. 
For each frequency channel  we consider an homogeneous detector noise with the power spectrum given by \cite{Perotto,Popa}:\\
\begin{eqnarray}
\label{noise1}
N^{c}_{l,\nu}=(\theta_b \Delta_a)^2 \exp^{ l(l+1) \theta^2_b / 8 \ln 2}
\hspace{0.3cm} c \in(T,P) \,,
\end{eqnarray}
where $\nu$ is the frequency of the channel, $\theta_b$ is the FWHM of the  beam and
$\Delta_c$ are the corresponding sensitivities per pixel.
The global noise of the experiment is obtained as:
\begin{equation}
\label{noise2}
N^c_l=\left[ \sum_{\nu} (N^{c}_{l,\nu})^{-1} \right]^{-1} \, .
\end{equation}
Our fiducial model is the standard $\Lambda$CMB cosmological model
with the  physical baryon  density $\Omega_{b}h^2=0.022$,
the physical dark matter density $\Omega_{cdm}h^2=0.11$,
the ratio of the sound horizon distance
to the angular diameter distance $\theta_s=1.04$,
the reionization optical depth $\tau= 0.085$,
the scalar spectral index $n_s=0.96$ and the curvature fluctuations amplitude
$A^2_S= 2.28 \times 10^{-9}$ at pivot scale k=0.01Mpc$^{-1}$\cite{Dunkeley09,Komatsu09}.\\
We present in Fig.~1 the CMB angular power spectra of the the fiducial $\Lambda$CMB cosmological model  and the
temperature and polarization noise power spectra obtained
for the {\sc Planck} experimental characteristics presented in Table~1 considering
a coverage of the sky of 80\%. 
\begin{figure}

\vspace{-3cm}
\epsscale{0.8}
\plotone{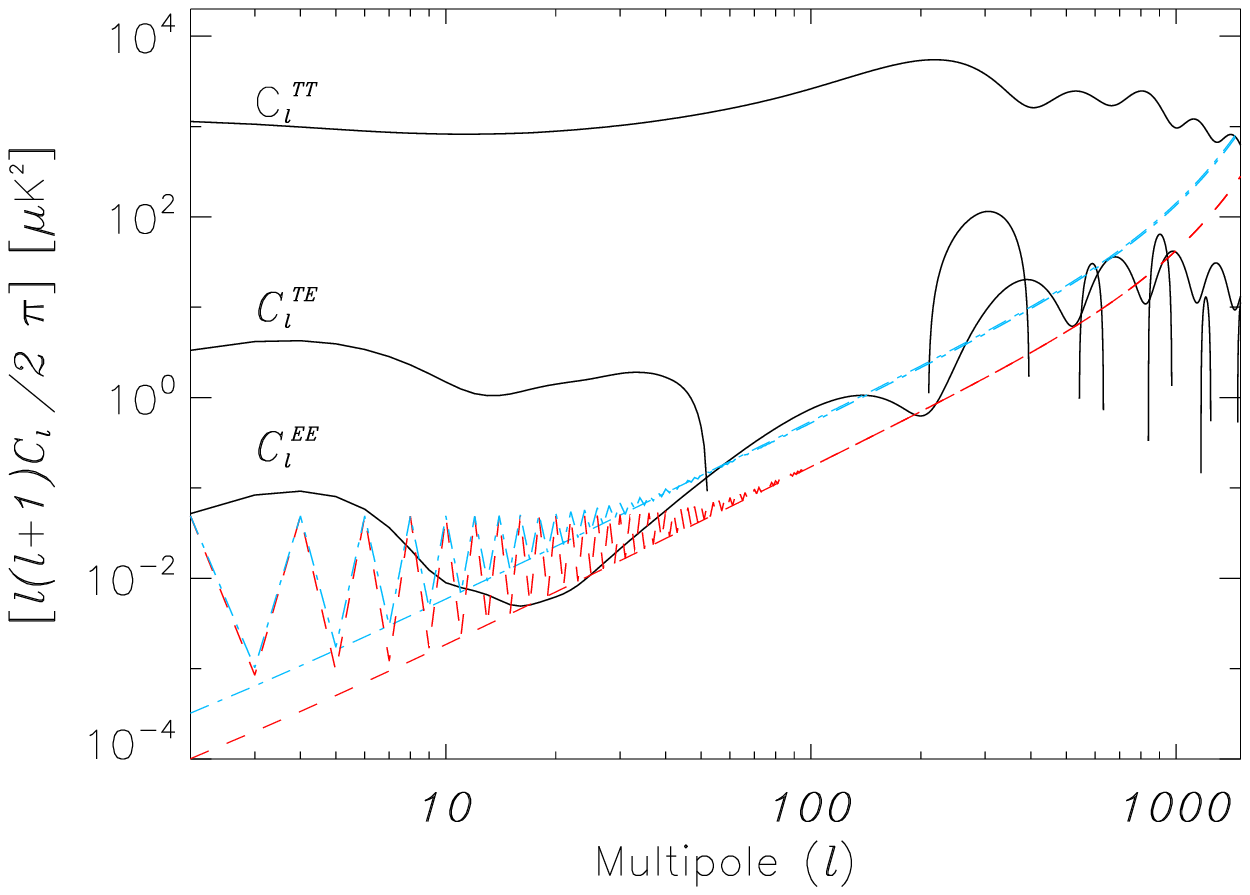}
\caption{The CMB angular power spectra of the fiducial $\Lambda$CMB cosmological model and the
temperature (dashed red line) and polarization (dash-dotted blue line)
noise power spectra obtained for the {\sc Planck} experimental characteristics presented in Table~1, considering a coverage of the sky of 80\%.}
\end{figure}
%======================\\
The final evaluation of the systematic effects
that could remain in the {\sc Planck} data after data reduction
affecting scientific exploitation
will come from accurate in-flight analyses and extensive
Monte Carlo simulations and is out of the scope of this work.
On the other hand, we include in this study also a degradation
of {\sc Planck} ideal sensitivity
possibly introduced by residuals of systematic effects
at low multipoles, where the cross-check for systematics possible
at high multipoles comparing different sky areas is
obviously not feasible.
The most critical source of contamination will likely come from
the straylight, e.g. the signal entering far sidelobes
\cite{sandri04} at large angular
distance from the main beam.
Two different sources mainly contribute to this effect: the CMB dipole
\cite{burigana_stray_dip} and the Galactic emission
\cite{burigana_stray_gal}. The former affects only even multipoles, but
it is larger in amplitudes at the considered frequencies,
the latter is smaller in amplitudes, but affect all multipoles.
Our simple conservative toy model, based on the above studies,
assumes an increasing of the noise power
at low multipoles coming from residuals of the
angular power spectra estimated for these systematic effects
possible generated by a non perfect subtraction of them, as in the
case in which the properties of {\sc Planck} optical response
in the far sidelobes is known only with an accuracy of
about 30\%. \\
The uncertainty added to the instrumental (receiver) noise
is clearly visible in Fig.~1.
%==================

We evaluate the likelihood function for 4-D and 5-D inflationary models by using
the public packages {\sc CosmoMC} and {\sc CAMB}
\cite{Lewis02,camb} modified to enable us to include the corresponding Hamilton-Jacobi formalism as described in the previous section. We perform the analysis in the framework of the flat $\Lambda$CDM
standard cosmological model.\\
For 4-D inflation case, the $\Lambda$CDM
standard cosmological model is described by the following sets of
parameters receiving uniform priors:
$$ \Omega_bh^2 \,,\,\,\Omega_ch^2\,,\,\, \theta_s\,,\,\, \tau\,,\,\,
A^2_S\,,\,\, \frac{H'^2}{H^2}\,,\,\, \frac{H''}{H}\,,\,\, \frac{H'''H'}{H^2}\,$$
where $H'$, $H''$ and $H'''$ are the derivatives of the Hubble expansion rate $H$
with respect to the scalar field. 
As notted before by Lesgourgues at al. (2008), because the  physical effects in the primordial power spectra depend on combinations of Hubble expansion rate derivatives,
the basis of parameters receiving uniform priors should consists in functions of the
above combinations or linear combinations of them, ensuring that
Markov Chains can converge in a reasonable amount of time. \\
By analogy, we take for 5-D inflation case the following basis of parameters receiving uniform priors:
$$  \Omega_bh^2 \,,\,\,\Omega_ch^2\,,\,\,  \theta_s\,,\,\, \tau\,,\,\,
A^2_S\,,\,\, \frac{y'^2}{y^2}\,,\,\, \frac{y''}{y}\,,\,\, \frac{y'''y'}{y^2} \,,\,\,\mu\,$$
where $y'$, $y''$ and $y'''$ are the derivatives with respect to the scalar field of
the parameter $y(\phi)$ defined by Eq.(\ref{yphi}).
One should note that the 5-D inflation case requires the additional parameter $\mu$
that controles the hierarchy of 4-D and 5-D Planck mass scales through
the brane tension $\lambda$:
\begin{equation}
\label{rapm}
\mu=\frac{m^3_5}{m_4} =
\sqrt{ \frac{4 \pi \lambda }{3} }\,.
\end{equation}
We run 32 Monte Carlo chains per model and dataset, imposing for each case the Gelman \& Rubin convergence criterion \cite{Gelman92}.

\section{The results: analysis and interpretation}
\subsection{The 4-D and 5-D inflationary parameter bounds}
\begin{figure}
\epsscale{1.}
\plotone{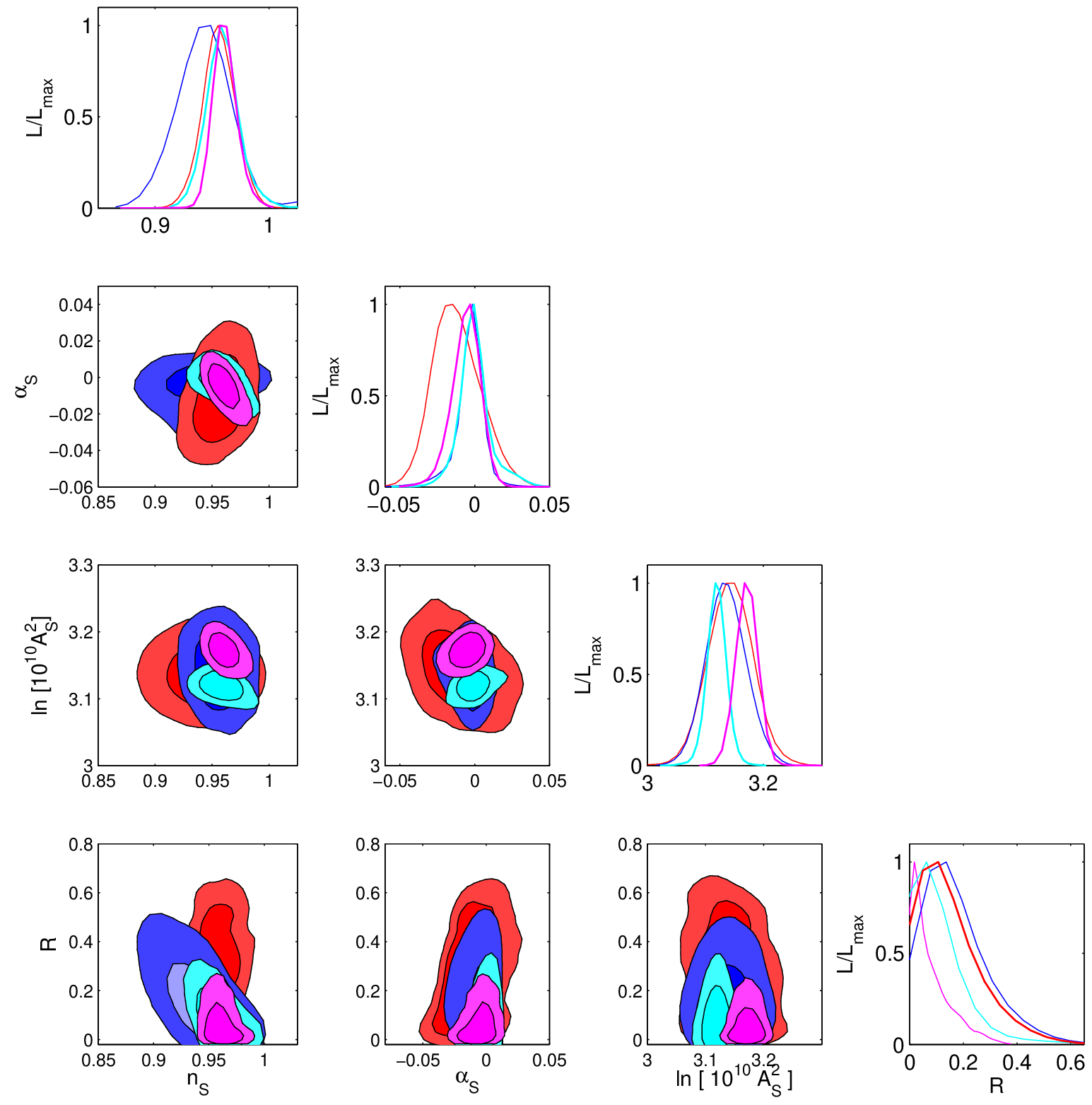}
\caption{We show the results of the fits to the WMAP5+SN+BAO dataset of
4-D inflation model in red and 5-D in inflation model in blue.
The results of the fits to the {\sc Planck}-like simulated dataset of
4-D inflation model are in magenta and of 5-D in inflation model are in cyan.
The top plot in each column shows the
probability distribution  of different scalar inflationary observables
while the other plots show their joint 68\% and 95\% confidence intervals.
All parameters are computed at the Hubble radius crossing $k_*$=0.01Mpc$^{-1}$.}
\end{figure}
\begin{figure}
\epsscale{0.8}
\vspace{-1cm}
\plotone{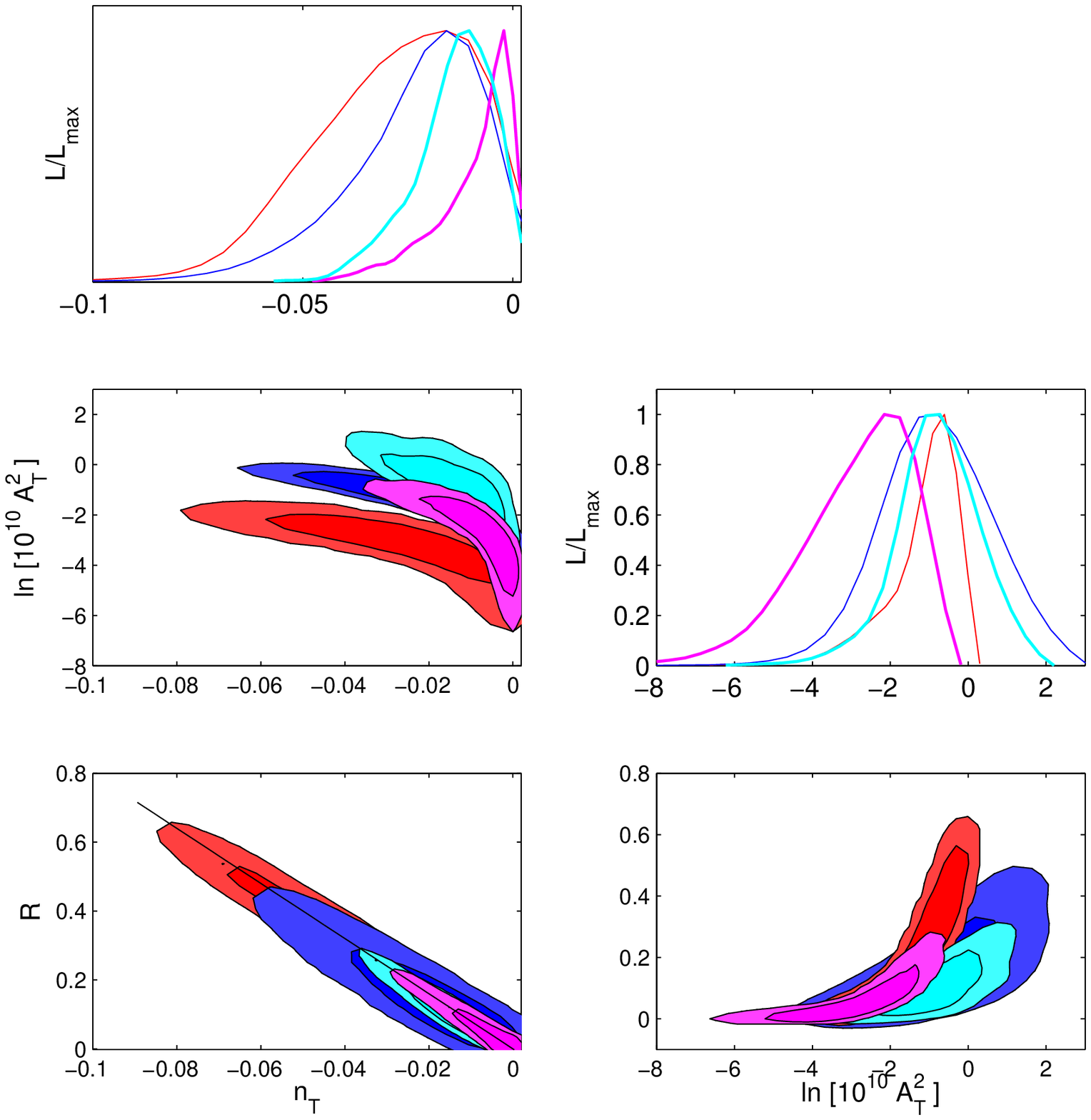}
\caption{We show the results of the fits to the WMAP5+SN+BAO dataset of
4-D inflation model in red and of 5-D in inflation model in blue.
The results of the fits to the {\sc Planck}-like simulated dataset of
4-D inflation model are in magenta and of 5-D in inflation model are in cyan.
The top plot in each column shows the
probability distribution  of different tensorial inflationary observables
while the other plots show their joint 68\% and 95\% confidence intervals.
All parameters are computed at the Hubble radius crossing $k_*$=0.01Mpc$^{-1}$.}
\end{figure}
The parameter bounds derived from each set of chains are given in Table~2 while
Fig.~2 and Fig.~3 show the results of our fits of 4-D and 5-D inflationary models
on WMAP5+SN+BAO dataset and {\sc Planck}-like simulated dataset.
All parameters are computed at the Hubble radius crossing $k_*$=0.01Mpc$^{-1}$. \\
From the fit of 4-D inflation model to WMAP5+SN+BAO dataset we obtain
bounds on $n_S$, $A^2_S$, $\alpha_S$ and R at
$k_*=$0.01 Mpc$^{-1}$ that translated into bounds at
$k_*=$0.002 Mpc$^{-1}$ show a good agreement with bounds on the similar parameters
 reported by the WMAP team \cite{Komatsu09}. 
Although our computation does not involve the HSR approximation 
in the computation of perturbations spectra, our constraints on the inflationary parameters obtained from the fit of 4-D inflation model to WMAP5+SN+BAO dataset are in general in agreement with the similar results
obtained by using the HSR formalism to recover the inflationary potential, when imposing constraints on the number of {\it e}-folds during inflation \cite{Peiris06a,Peiris06b}. 
Our results are directly comparable with
the results presented in Hamann et al. (2008) that uses
the same numerical evaluation of the spectra to obtain constraints on the inflationary parameter
by using a selection of CMB data including WMAP complemented by LSS measurements.\\
%&&&&&&&&&&&&&&&&&&&&&&&&&&&&&&&&&&&&&&&&&&&&&&&&&&&&&&&&&&&&&&&&&&&&
Looking at Fig.~2 and Fig.~3  we see that
the amplitude of scalar power spectrum $A^2_S$ obtained
from both datasets is suppressed in the 5-D inflation case,
when compared with the similar values obtained in 4-D standard inflation.  
Morever, in  5-D inflation model the joint confidence regions of the scalar spectral index $n_s$ and tensor to scalar amplitude ratio $R$
are anti-correlated, while the amplitude of the tensor power spectrum $A^2_T$ is increased,
when compared with the 4-D standard inflation case. 
This  can be attributed to a larger contribution of the
tensor modes to the primordial perturbations in braneworld inflation.
In this case the normalization of the scalar perturbation
is reduced to obtain the correct value of $R$. 
Likewise, we obtain  a strong correlation between the values of $R$ and the tensor spectral index
$n_T$ from the fits of 4-D and 5-D inflation models to both datasets, in good agreement with the predictions of consistency relation given in Eq.~(1). \\
\begin{figure}
\epsscale{1.}
\plotone{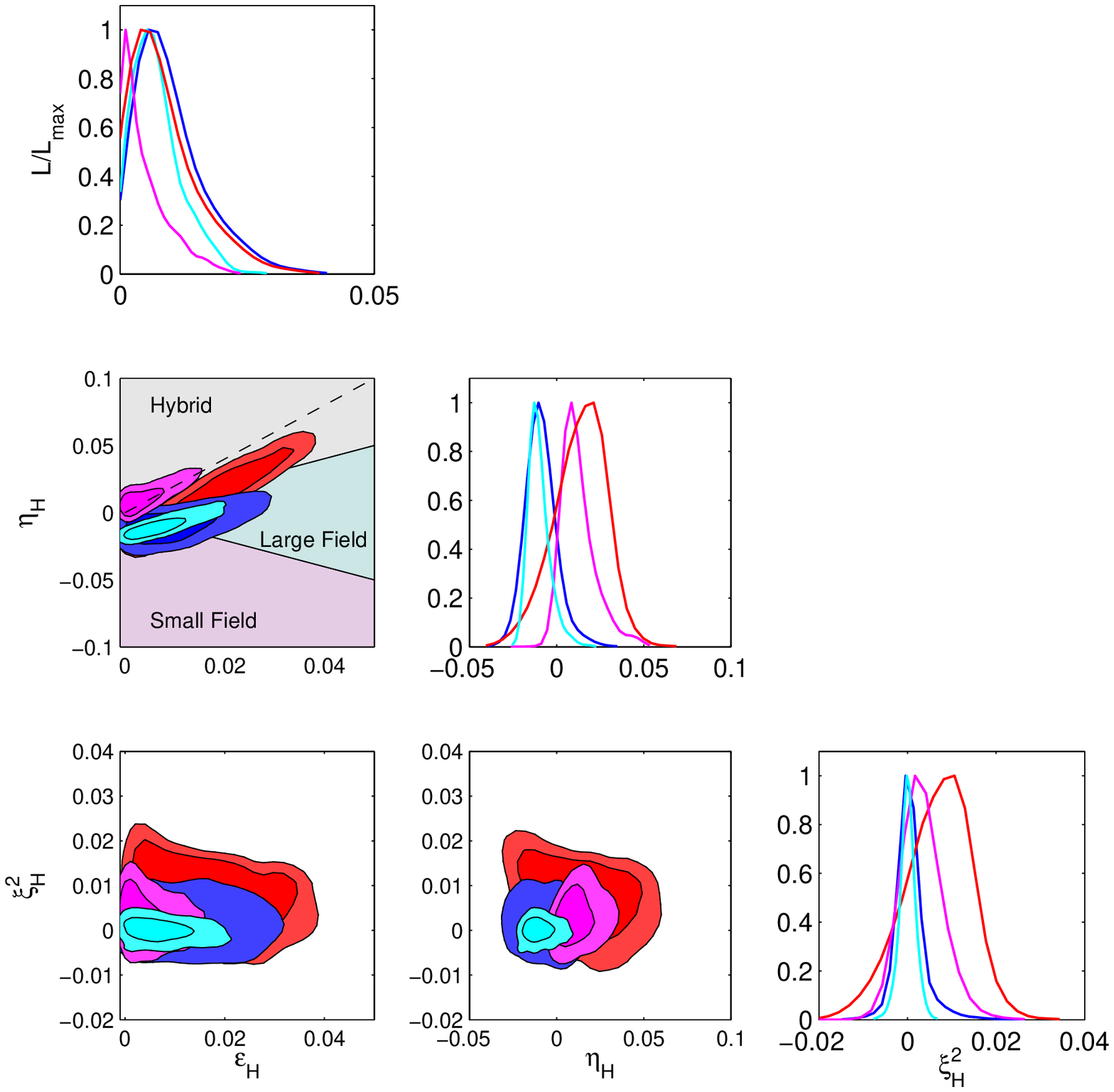}
\caption{We show the bounds on the HSR parameters derived from the fits to
WMAP5+SN+BAO dataset of 4-D inflation model in red and of 5-D inflation model in blue.
The results of the fits to the {\sc Planck}-like simulated dataset of
4-D inflation model are in magenta and of 5-D in inflation model are in cyan.
The top plot in each column shows the
probability distribution  of different HSR parameters
while the other plots show their joint 68\% and 95\% confidence intervals.
All parameters are computed at the Hubble radius crossing $k_*$=0.01Mpc$^{-1}$.
We show the division of $\epsilon_H-\eta_H$ plane into
large field ($-\epsilon_H < \eta_H < \epsilon_H$), small field  ($\eta_H < -\epsilon_H$) and hybrid field  ($0  < \epsilon_H < \eta_H$) classes of inflation models.}
\end{figure}
Fig.~4 presents the constraints on HSR
parameters derived from the fits of 4-D and 5-D inflationary models
on WMAP5+SN+BAO dataset and {\sc Planck}-like simulated dataset.
We show the division of $\epsilon_H$-$\eta_H$ plane into
large field, small field and hybrid field classes of inflation models
\cite{Kinney02,LidT02} overlaid with our constraints on
their joint  68\% and 95\% confidence intervals. 
We see that all three classes of inflation models are allowed at 2-$\sigma$ level
by the fit of 4-D standard inflation to our datasets. \\
The joint marginalized distribution of $\epsilon_H$ and $\eta_H$
obtained from the fit of  5-D inflation model
shows that the large field and small field
classes of inflationary models are allowed by WMAP5+SN+BAO and by {\sc Planck} datasets at 2-$\sigma$ level, while the hybrid class of inflationary models seems to be disfavored by both datasets in the 5-D single-field inflation scenario.
The parameter values within each class of allowed inflationary models are tightly constrained by {\sc Planck} dataset. \\
Of particular interest are the differences between the degeneracy directions in $\epsilon_H$-$\eta_H$ plane found from the fit of 4-D inflation model to WMAP5+SN+BAO dataset and {\sc Planck} dataset that arise due to the dependence of $\alpha_s$ on $\xi_H^2$.
The role of the $\xi_H^2$  in the dynamics of inflation is discussed in details in \cite{Efstat05,Easther06} and the accuracy
of  slow-roll inflation models with signiﬁcant running 
is probed by using Monte Carlo reconstruction
in \cite{Easther03,Makarov05,Peiris06b}.\\
Looking at Fig.~4  one can see the preference of WMAP5+SN+BAO dataset
for large and positive $\xi^2$ values in 4-D standard inflation case,
that translates into large negative values of the running of scalar spectral index
 $\alpha_s$  and  a larger degeneracy in $n_S$-$\alpha_S$ plane,
 when compared to the similar results obtained from the fit to {\sc Planck} dataset (see Fig.~2).\\
The differences between the degeneracy directions obtained in 4-D standard inflation case and 5-D inflation arise via the dependence of HSR parameters on the
dynamical equations driving inflation, which are different in 4-D and 5-D inflationary models.

\subsection{Reconstruction of 4-D and 5-D inflationary potential}

The aim in the reconstruction of the inflationary potential is to take the measurements of various inflationary observables
corresponding to a particular wavenumber $k$
and to use them to obtain the inflationary potential $V(\phi)$ and its
derivatives at the scalar field value $\phi_*$ when the scale $k$ crosses the Hubble radius $k_*$ during inflation. \\
In the general case of the single-field braneworld inflation,
the slope and the curvature of the 5-D inflationary potential
as function of inflationary observables $R$ and $n_s$ and on the combination $V/\lambda$ are given by \cite{LidT02}:
\begin{eqnarray}
\frac{V'}{V} & = & \sqrt{\frac{16 \pi R}{m^2_4}}
            \left[\frac{1+V/2\lambda}{G(V/\lambda)}\right] \,,\\
 \frac{V''}{V} & = & \frac{4 \pi}{m^2_4}\left(1+\frac{V}{2 \lambda}\right)
\left[6R \,\frac{1+V/\lambda}{G^2(V\lambda)} + (n_S-1) \right]\,,
\end{eqnarray}
where:
\begin{eqnarray}
G(x) & = & \left[ \sqrt{1+x^2} -x^2 {\rm sinh}^{-1}  \frac{1}{x}  \right]^{-1/2}\,.
\end{eqnarray}
In the high-energy limit ($V >> \lambda$) the function
$G^2(V/\lambda) \rightarrow 3V/2 \lambda$.
In the low-energy limit ($\lambda >> V$) $G^2(V/\lambda) \rightarrow 1$
and the scalar and tensor perturbation spectra of 4-D standard inflation are recovered.\\
%-----------------------------------
\begin{figure}
\epsscale{1.}
\plotone{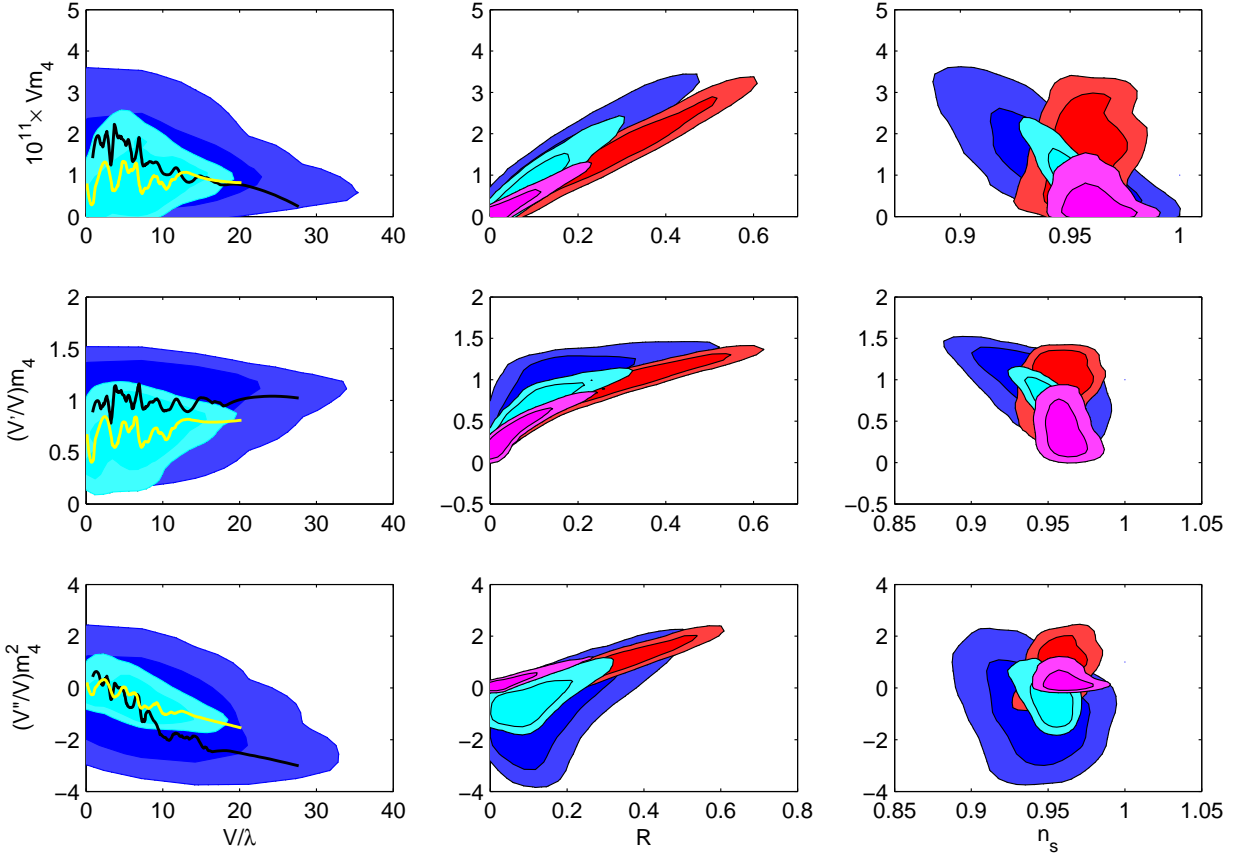}
\caption{We presesent the recovered regions (at 68\% and 95\% CL) of the magnitude, slope and curvature of the inflationary potential from the fits to
WMAP5+SN+BAO dataset of 4-D inflation model in red and of 5-D
inflation model  in blue.
The similar results from the fits to {\sc Planck}-like simulated dataset of
4-D inflation model are in magenta and of 5-D in inflation model are in cyan.
We also show the 1D marginal distribution of the recovered magnitude of the
inflationary potential, its slope and curvature (at 95\% CL)
as function of $V/\lambda$ obtained from the fit of 5-D inflation model
to WMAP5+SN+BAO dataset (black lines) and {\sc Planck}-like simulated dataset
(yellow lines). All parameters are computed at the Hubble radius crossing $k_*$=0.01Mpc$^{-1}$.}
\end{figure}
%-----------------------------------------------
\begin{figure}
\epsscale{0.8}
\plotone{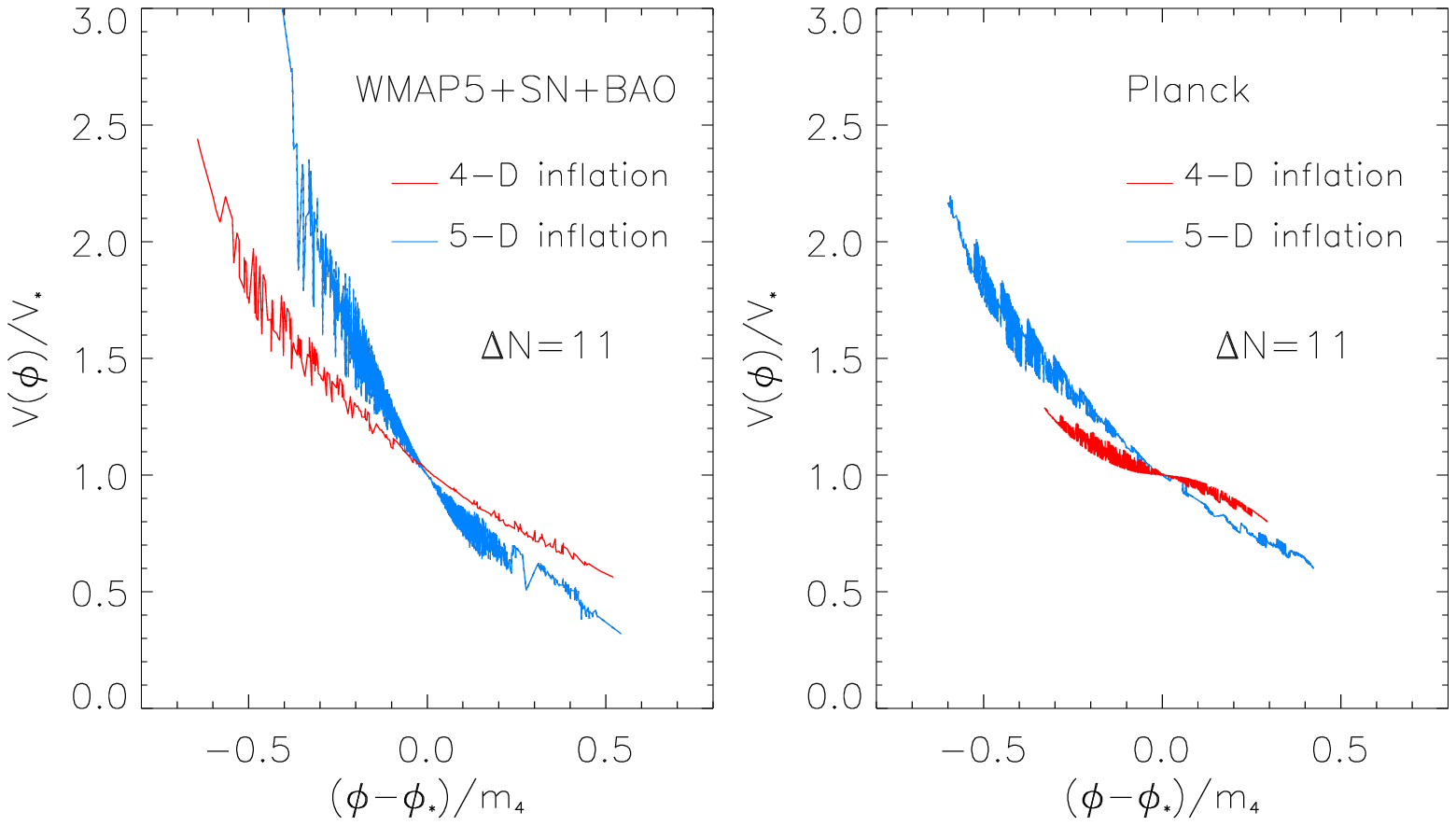}
\caption{We present the regions of 4-D and 5-D inflationary potentials
allowed at 65\% CL by WMAP5+SN+BAO and {\sc Planck}-like simulated datasets
in the same observational inflation window corresponding to $\Delta N =11$ {\it e}-folds.
For each case, the magnitude of inflationary potential is normalized to $V_*$, the value of the inflationary potential at Hubble radius crossing $k_*$=0.01Mpc$^{-1}$.}
\end{figure}
We compute the magnitude, the slope and the curvature of the inflationary potential from the fits of 4-D and 5-D inflation models to
our datasets by using  Eq.(10) and Eq.(18) respectively.
In Fig.~5 we show the allowed  regions of the recovered magnitude of the inflationary potential,
its slope and curvature from the fit of 4-D and 5-D inflation models to
WMAP5+SN+BAO dataset and {\sc Planck}-like simulated  dataset.
We also show  the 1D marginal distribution of the recovered inflationary potential,
its slope and curvature as a function of $V/\lambda$,  obtained  from the fit of 5-D inflation model to the same datasets. \\
Fig.~5 explicitly demonstrates the effect of the braneworld  reconstruction
of the inflationary potential. 
As $V/\lambda$ is increased
the magnitude and the curvature of the inflationary potential are decreased while its slope steepens.
Also, the magnitude,  the slope and the curvature of the inflationary potential are increased
in both 4-D and 5-D inflationary scenatios when $R$ increases.\\
The results from the fit of 5-D inflation model to WMAP5+SN+BAO dataset
and {\sc Planck}-like simulated  dataset show 
that the magnitude and the slope of the inflationary potential
are anti-correlated with the scalar spectral index, $n_s$.\\
%show  an anti-correlation between
%the magnitude of the inflationary potential, its slope and the scalar %spectral index, $n_s$.\\
The mean values of the magnitude, slope and curvature of the inflationary potentials together with their 95\% upper and lower intervals are given in Table~2.\\
In Fig.~6 we show the dependence of the reconstructed regions of 4-D and 5-D inflationary
potentials allowed  by the same datasets (at 65\% CL) in an observational
inflationary window corresponding to $\Delta N =11$ {\it e}-folds, as functions of the scalar field.

\subsection{The 4-D and 5-D single-field inflation consistency relations}
\begin{figure}
\epsscale{1.}
\plotone{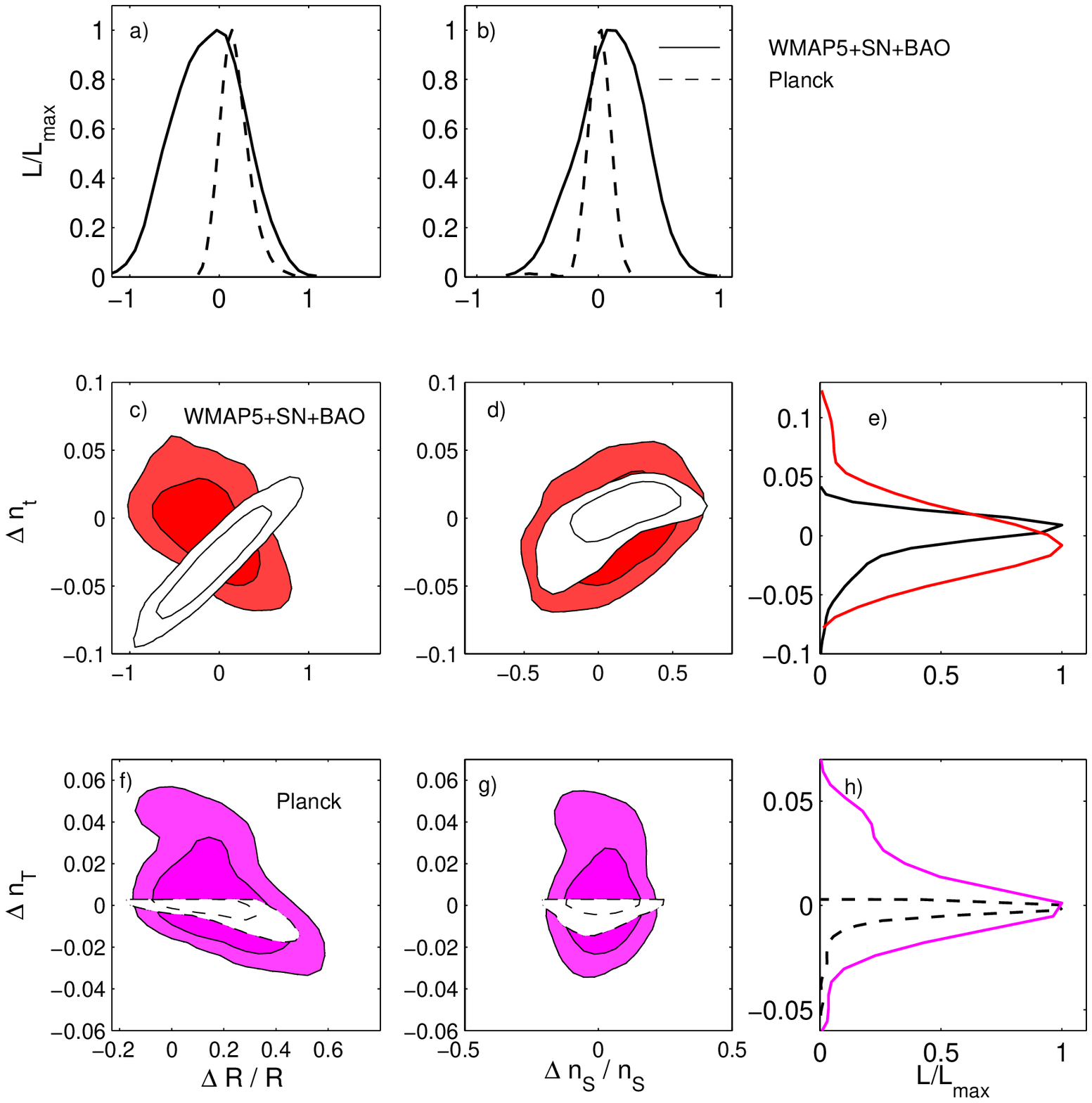}
\caption{Panels a) and b): 1D marginalized likelihood probability distributions of the theoretical errors
$\Delta R/R$ and  $\Delta n_S/n_S$  obtained from the fits of 4-D standard inflation model to WMAP5+SN+BAO dataset (continuous line)
and {\sc Plaanck} dataset (dashed line) assuming $\Delta N$=14.
Panels c) and d): 2D joint marginalized probability distributions (at 68\% and 95\% CL) obtained from the fits to WMAP4+SN+BAO dataset.
The red filled contours corresponds to
 $\Delta n_T =\Delta n^{4D}_T- \Delta n^{5D}_T$ and the white filled contours
corresponds to the theoretical error $\Delta n_T^c$ obtained by assuming $\Delta N$=14. Panel e):  1D marginalized likelihood probability distributions of
$\Delta n^c_T$ (black continuous line) and  $\Delta n_T=n^{4D}_T-n^{5D}_{T}$ (red continuous line) obtained from the fits to WMAP5+SN+BAO dataset.
Panels e) and g):  2D joint marginalized probability distributions (at 68\% and 95\% CL) obtained from the fits to {\sc Planck}-like simulated  dataset. The magenta filled contours corresponds to
 $\Delta n_T =\Delta n^{4D}_T- \Delta n^{5D}_T$ and the white filled contours
corresponds to the theoretical error $\Delta n_T^c$ obtained by assuming
$\Delta N$=14. Panel h):  1D marginalized likelihood probability distributions of
$\Delta n^c_T$ (black dashed line) and  $\Delta n_T=n^{4D}_T-n^{5D}_{T}$ (magenta continuous line) obtained from the
fits to {\sc Planck}  dataset.}
\end{figure}
There is an infinite hierarchy of consistency equations
of the single-field standard inflation 
\cite{Lidsey97,Song03,Chung03,Chung06,Cortes06}. 
To the leading order in the slow-roll approximation,
the consistency relation of the standard scenario given in Eq.(1) is degenerate. 
To next-to-leading order, this consistency relation receives
corrections of the form \cite{Copeland94,Lidsey97}:
\begin{equation}
\label{consist2}
n_T=-2\frac{A^2_T}{A^2_S}\left[1-\frac{A^2_T}{A^2_S}+(1-n_S)\right]\,,
\end{equation}
that do not depend on the spectral index of the tensor perturbations.\\
As the inflationary observables $n_S$, $n_T$ and $R$ are evaluated at
the epoch of horizon-crossing quantified by the number of {\it e}-folds $N$
before the end of the inflation at which our present Hubble scale
equalled the Hubble scale during inflation, the uncertainties in the determination of $N$
translates to theoretical errors in the determination of the inflationary observables. Assuming that the ratio of the entropy per comoving interval today to that after reheating is negligible,
the main uncertainty in the determination of $N$  is caused by our ignorance in the determination of the reheating temperature after inflation leading to an error of $\Delta N \sim 14$ 
\cite{Kinney06,Adshead08}.\\
%&&&&&&&&&&&&&&&&&&&&&&&&&&&&&&&&&&&&AICIIIIIIIIIIIIIIII
In order to test the observational signature that standard and
braneworld inflationary scenarios may produce, 
we use the estimates of the inflationary parameters 
obtained from the fits to  WMAP5+SN+BAO and 
{\sc Planck}-like simulated datasets to compare the experimental 
difference between tensor spectral indexes, $n^{4D}_T-n^{5D}_T$, 
to the theoretical error in the tensor spectral index 
computed by using the consistency relation (\ref{consist2}).\\ 
To the lowest order in slow-roll parameters, 
the  uncertainties $\Delta R$ and $\Delta n_S$ in terms of the uncertainty in
the number of {\it e}-folds $\Delta N$ are given by \cite{Kinney02,Kinney04}:
\begin{eqnarray}
\frac{\Delta R}{\Delta N}=R\left[(n_S-1)+ \frac{R}{8}\right] \, ,\\
\frac{\Delta n_s}{\Delta N} = -\frac{5}{16}R(n_S-1)-\frac{3}{32}R^2+2 \xi^2 \,.
\end{eqnarray}
The theoretical uncertainty on the tensor spectral index in the standard 4-D inflation can be straightforward obtained from Eq.(\ref{consist2}) by using Eqs.(35) and (36):
\begin{eqnarray}
%\frac{\Delta n^c_T}{\Delta N} & = &-\frac{1}{4}\frac{\Delta R}{\Delta N}+ %\frac{1}{4}\frac{A^2_T}{A^2_S}\frac{\Delta R}{\Delta N}
%+\frac{1}{8} \frac{A^2_T}{A^2_S}\frac{\Delta n_S}{\Delta N}+ \frac{n_S}{8}\frac{\Delta R}{\Delta N} \,.\\
\frac{\Delta n^c_T}{\Delta N} & = &\frac{1}{4}\left[1-\frac{n_S}{2}-\frac{A^2_T}{A^2_S}\right]\frac{\Delta R}{\Delta N}+\frac{1}{8} \frac{A^2_T}{A^2_S}\frac{\Delta n_S}{\Delta N}\,.
\end{eqnarray}
The estimate of $\Delta n^c_T/\Delta N$ should be compared to 
$\Delta n_T=n$. In Table~3 we present the mean values of the 
lowest order estimates of the theoretical errors $\Delta R/R$, $\Delta n_S/n_S$ and  $\Delta n^c_T$
from the fit of 4-D inflation model to WMAP5+SN+BAO and {\sc Planck} datasets 
obtained by assuming $\Delta N$=14
{\it e}-folds and the mean values of the difference 
between the experimental values of the tensor spectral 
indexes $\Delta n_T=n^{4D}_T-n^{5D}_T$, while
in Fig.~7 we show their 1D marginalized likelihood probability distributions.
We also show in the same figure the 1D marginalized likelihood probability 
distribution of the lowest order estimates of the  theoretical errors
$\Delta n^c_T$, obtained by assuming $\Delta N$=14, compared with 
the 1D marginalized likelihood probability distribution of
the difference $\Delta n_T=n^{4D}_T-n^{5D}_T$ and their 
2D joint allowed bounds (at 68\% and 95\% CL). \\
The analysis of the results presented in  Fig.~7 and Table 3 shows that
the $\Delta n_T$ parameter space obtained from the fits of 4-D and 5-D inflation models to WMAP5+SN+BAO dataset is dominated by the theoretical error $\Delta n^c_T$: the confidence
interval corresponding to $\Delta n^c_T$ is smaller by a factor of
1.2 than that corresponding to $\Delta n_T$. The same parameter space is better 
constrained  by the {\sc Planck} dataset: in this case
the confidence interval corresponding to $\Delta n^c_T$ is
three times smaller than that corresponding to $\Delta n_T$. \\
We conclude that the detection of tensor perturbations and
the theoretical uncertainties in the inflationary observable represent
a significant challenge for the future {\sc Planck} CMB measurements:
distinguishing  between the observational signatures of the standard and
braneworld single-field inflation scenarios.

\section{Conclusions}

One of the most anticipated results of forthcoming {\sc Planck} 
high precision CMB measurements is probing the physics of inflation 
and in particular, the reconstruction of the inflation potential.
On the other hand, the possibility that our four-dimensional Universe could 
lie on a brane embedded in a higher dimensional space has important consequences 
for the early universe  and in particular for the cosmological inflation. 

In this paper we make a more general determination of the inflationary observables
in the 4-D and 5-D single-field inflationary scenarios
by exact reconstruction of the dynamics of the inflation potential 
during the observable inflation, with a minimal number of assumptions. \\
Making use of the general formalism for 5-D single-field inflation 
developed by Hawkins and Lisdey \cite{HawL01,HawL03} valid in all regimes, 
having many of the properties of the {\em Hamilton-Jacobi}
formalism developed for the 4-D standard inflation, 
we compute the scale dependence of the amplitudes of the scalar
and tensor perturbations by integrating the exact mode equation. 
Our computation does not assume the slow-roll approximation and 
is valid in all regimes if the field is monotonically rolling 
down its potential.
The solutions in 4-D and 5-D inflation scenarios differ through 
the dynamics of the background scalar field and the number
of {\it e}-folds assumed to be compatible with the observational 
window of inflation. \\
We address higher-order effects in the standard and braneworld single-field
inflation scenarios by fitting the Hubbble expansion rate $H(\phi)$ 
and subsequently the inflationary potential $V(\phi)$, directly
to WMAP5+SN+BAO  and Planck-like simulated datasets. \\
One should note that our results refer to the initial scalar 
and tensor perturbation spectra and
not to the braneworld effects on the subsequent evolution 
of the perturbations that is likely to
be model dependent \cite{Challinor,Brax}. \\
Assuming that the ratio of the entropy per comoving interval today to that after reheating is negligible, we analyze the implications of the theoretical uncertainty in the determination of the reheating temperature after inflation on the observable predictions of inflation.\\
We find that the detection of tensor perturbations and
the theoretical uncertainties in the inflationary observables represent
a significant challenge for the future {\sc Planck} CMB measurements:
distinguishing  between the observational signatures of the standard and
braneworld single-field inflation scenarios.

\begin{deluxetable}{cccc}
\tablewidth{0pt}
\tablecaption{The expected  experimental characteristics
for the {\sc Planck} frequency channels considered in the paper \cite{Mandolesi_etal_M1,Blue}.
$\Delta_T$ and $\Delta_P$ are the sensitivities per pixel for temperature
and polarization maps.}
\tablehead{Frequency $(\nu)$ & FWHM  & $\Delta_T  $ & $\Delta_P $ \\
(GHz) &(arc-minutes) & ($\mu$K)& ($\mu$K)  }
\startdata
70 & 13  & 23.48 & 33.21  \\
100 & 9.5 &6.8 & 10.9 \\
143 & 7.1 &6.0 & 11.4 \\
\enddata
\end{deluxetable}
\begin{deluxetable}{lcccc}
%\rotate
\tablewidth{0pt}
\tablecaption{The mean values and 95\% CL lower and upper intervals
of the derived parameters obtained from the fit of 4-D and 5-D inflation models
to WMAP5+SN+BAO dataset and {\sc Panck}-like simulated dataset.
All parameters are computed at the Hubble radius crossing $k_*$=0.01 Mpc$^{-1}$.}
\tablehead{  & \multicolumn{2}{c}{WMAP5+SN+BAO} & \multicolumn{2}{c}{{\sc Planck}} \\
Parameter& 4-D Inflation & 5-D Inflation & 4-D Inflation & 5-D Inflation
}
\startdata

$\Omega_bh^2$& $0.022_{0.021}^{0.023}$  & $0.022_{0.021}^{0.023}$
             & $0.022_{0.021}^{0.023}$  & $0.022_{0.021}^{0.023}$   \\
$\Omega_ch^2$& $0.111_{0.106}^{0.117}$  &  $0.111 _{0.106}^{0.117}$
             & $0.113_{0.112}^{0.122}$  & $0.112_{0.111}^{0.123}$     \\
$\tau$       & $0.082_{0.055}^{0.109}$  & $0.080_{0.056}^{0.106}$
             & $0.081_{0.734}^{0.897}$  & $0.081_{0.073}^{0.903}$ \\
$\theta_s$   & $1.039_{1.034}^{1.045}$  & $1.039_{1.034}^{1.044}$
             & $1.050_{1.048}^{1.051}$  & $1.053_{1.051}^{1.053}$  \\
${\rm ln}[ 10^{10}A^2_S ]$  & $3.143_{3.083}^{3.201}$& $3.061_{2.986}^{3.137}$
                            & $3.172_{3.146}^{3.194}$& $3.103_{3.077}^{3.142}$
    \\
$\epsilon_H $&  $< 0.035$ & $ <0.024$ & $ <0.019$&   $ <0.017 $\\
$\eta_H$ & $0.011_{-0.022}^{\,\,\,\,0.051}$&  $-0.008_{-0.022}^{\,\,\,\,0.011}$&
$-0.006_{-0.022}^{\,\,\,\,0.013}$ & $-0.009_{-0.019}^{\,\,\,\,0.011}$\\
$\xi^2$ & $0.007_{-0.006}^{\,\,\,\,0.017}$ & $0.001_{-0.003}^{\,\,\,\,0.011}$& $-0.001_{-0.006}^{\,\,\,\,0.011}$
&$0.001_{-0.007}^{\,\,\,\,0.008}$  \\
%&&&&&&&&&&&&&&&&&&&&&&&&&&&&&&&&&&&&&&&&&&&&&&&&&&&
$n_S$ & $0.956_{0.932}^{0.979} $  & $0.947_{0.906}^{0.995}$
      & $0.958_{0.948}^{0.964} $  & $0.968_{0.940}^{1.058}$\\
$\alpha_S$  & $-0.012_{-0.038}^{\,\,\,\,0.021}$
                   & $-0.006_{-0.021}^{\,\,\,\,0.007}$
                   & $-0.005_{-0.022}^{\,\,\,\,0.006}$ & $0.000_{-0.015}^{\,\,\,\,0.021} $\\
$n_T$& $ > -0.042$ &  $ > -0.055 $
                   & $ > -0.028 $ & $ > -0.043$\\
${\rm ln}[ 10^{10}A^2_T ]$ & $-1.174_{-3.071}^{-0.203}$  & $-0.836_{-3.095}^{\,\,\,\,1.477}$
                           & $-2.867_{-5.501}^{-1.042}$  & $-0.448_{-2.983}^{\,\,\,\,2.018}$ \\
$R$ & $ < 0.556$ &  $ <  0.476$
    & $ < 0.278$& $< 0.220$  \\
\hline
$V/\lambda$ &     -& $8.131_{\,\,\,\,0.002}^{20.632}$
                    &     -&$3.519_{0.002}^{6.383}$    \\
$10^{11}\times V m_4$ & $1.511_{0.174}^{3.004}$&$1.342_{0.244}^{2.783}$
&$0.449_{0.001}^{1.434}$&$0.933_{0.088}^{2.131}$      \\
$(V'/V)m^2_4$      & $0.895_{0.329}^{1.301}$& $0.922_{0.402}^{1.307}$
&$0.423_{0.090}^{0.871}$& $0.680_{0.222}^{1.057}$     \\
$(V''/V)m^2_4$      & $0.619_{-0.848}^{\,\,\,\,2.051}$&$-0.683_{-0.265}^{\,\,\,\,1.195}$&
            $0.356_{-0.014}^{\,\,\,\,1.116}$&$-0.479_{-1.437}^{\,\,\,\,0.774}$      
\enddata
\end{deluxetable}
\begin{deluxetable}{ccc}
\tablewidth{0pt}
\tablecaption{The mean values and 95\% CL lower and upper intervals of the lowest order estimates of
the theoretical errors $\Delta R/R$, $\Delta n_S/n_S$ and  $\Delta n^c_T$
obtained from the fit of 4-D inflation model to WMAP5+SN+BAO and {\sc Planck} datasets  by assuming $\Delta N$=14,
 and the mean values and 95\% CL lower and upper intervals of $\Delta n_T=n^{4D}_T - n^{5D}_T$,
obtained from the fits of 4-D and 5-D inflation models to the same datasets.}
\tablehead{   & WMAP5+SN+BAO &  {\sc Planck}
}
\startdata
$\Delta R/R$ & $-0.115_{-0.727}^{\,\,\,\,0.493}$ &  $\,\,\,\,0.177_{-0.061}^{\,\,\,\,0.478}$\\
$\Delta n_S /n_S$&\,\,\,\,$ 0.104_{-0.352}^{\,\,\,\,0.512}$& $-0.006_{-0.140}^{\,\,\,\,0.140}$ \\
$\Delta n^C_T$ & $-0.002_{-0.045}^{\,\,\,\,0.032} $& $-0.004_{-0.022}^{\,\,\,\,0.000}$\\
$\Delta n_T=n^{4D}_T-n^{5D}_T$& $-0.003_{-0.048}^{\,\,\,\,0.049}$& $\,\,\,\,0.003_{-0.026}^{\,\,\,\,0.043}$
\enddata
\end{deluxetable}

\acknowledgments
We acknowledge the use of the GRID computing system facility at the Institute
for Space Sciences Bucharest and to the staff working there.\\
L.P and A.C. are partially supported by ESA/PECS Contract C98051
and CNCSIS Contract 539/2009.

\end{document}